\documentclass[fleqn]{aa}  
\pdfoutput=1
\usepackage{graphicx}
\usepackage{graphicx,color,units}
\usepackage[fleqn]{amsmath} 
\usepackage[varg]{txfonts}
\usepackage{natbib,xspace}

\usepackage[switch]{lineno}
\newcommand{\const}{\ensuremath{\mathrm{const.}}}
\renewcommand{\div}{\ensuremath{\vec{\nabla} \cdot}}
\newcommand{\asr}{Adv. Space Res.}

\renewcommand{\deg}{\ensuremath{^{o}}}
\newcommand{\lamC}{$\lambda$ Cephei\xspace}

\begin{document}

\title{Cosmic rays in astrospheres}
\author{K. Scherer\inst{1,2} \and A. van der Schyff \inst{3} \and
  D.J. Bomans\inst{4,2} \and S.E.S. Ferreira\inst{3} \and
  H. Fichtner\inst{1,2} \and J. Kleimann\inst{1} \and
  R.D. Strauss\inst{3} \and K. Weis\inst{4} \and
  T. Wiengarten\inst{1} \and T. Wodzinski\inst{4}}
\institute{ Institut f\"ur Theoretische Physik IV: Weltraum- und
  Astrophysik, Ruhr-Universit\"at Bochum, Germany,
  \email{kls@tp4.rub.de, hf@tp4.rub.de, jk@tp4.rub.de, tow@tp4.rub.de}
\and  Research Department, Plasmas with Complex Interactions, Ruhr-Universit\"at Bochum, Germany  
  \and Center for Space Research, North-West University, 2520
  Potchefstroom, South Africa,
  \email{12834858@nwu.ac.za, Stefan.Ferreira@nwu.ac.za,
  DuToit.Strauss@nwu.ac.za} 
\and Astronomische Institute, Ruhr-Universit\"at Bochum, Germany,
  \email{bomans@astro.rub.de, kweis@astro.rub.de,
    thomas.wodzinski@astro.rub.de} 
} 
\date{Received; accepted}

\abstract
{Cosmic rays passing through large astrospheres can be efficiently
  cooled inside these ``cavities'' in the interstellar medium. Moreover,
  the energy spectra of these energetic particles are already
  modulated in front of the astrospherical bow shocks.}
{We study the cosmic ray flux in and around \lamC as an
  example for an astrosphere. The large-scale plasma flow is modeled
   hydrodynamically with radiative cooling. }
 {We studied the cosmic ray flux in a stellar wind cavity using a
   transport model based on stochastic differential equations. The
   required parameters, most importantly, the elements of the diffusion
   tensor, are based on the heliospheric parameters. The magnetic field
   required for the diffusion coefficients is calculated
   kinematically. We discuss the transport in an 
   astrospheric scenario with varying parameters for the transport
   coefficients. }
{We show that large stellar wind cavities can act as sinks for the
  galactic cosmic ray flux and thus can give rise to small-scale  anisotropies in
  the direction to the observer.  }
{ Small-scale cosmic ray anisotropies can naturally be explained by the modulation of
  cosmic ray spectra in
   huge stellar wind cavities.}
\keywords{Stars: winds, outflows -- Hydrodynamics -- ISM: cosmic rays}

   \maketitle

\section{Introduction}

Recently, simulations of astrospheres around hot stars have gained new
interest, see for example \citet{Decin-etal-2012},
\citet{Cox-etal-2012}, \citet{Arthur-2012},
\citet{van-Marle-etal-2014}. These authors modeled astrospheres using
a (magneto-)hydrodynamic approach, either in 1D or 2D. In this work,
such astrospheric models are used for the first time to estimate the
cosmic ray flux (CRF) through it. Because of the large spatial extent
of O~star astrospheres (wind bubbles), these objects can efficiently
cool the spectrum of galactic cosmic rays (GCR). Especially \lamC is
an interesting example, being the brightest runaway Of star in the
sky (type O6If(n)p). We estimate the CRF at different energies for
\lamC as an example of an O star astrosphere using stochastic
differential equations (SDE) \citep{Strauss-etal-2013} to solve the
GCR transport equation.  Runaway O and B stars are common and part
of a sizable population in the galaxy, a significant number of
which show bow-shock nebulae \citep[e.g.,][]{Huthoff-Kaper-2002}.  In
Sect.~\ref{sec:1} we discuss the radiative cooling functions, while
in Sect.~\ref{sec:2} we show the astrosphere model results. In Sect.~\ref{sec:3} we estimate the CRF.

\section{Large-scale structure of astrospheres}

Winds around runaway stars, or in general, stars with a nonzero
relative speed with respect to the surrounding interstellar medium
(ISM), develop bullet-shaped astrospheres.

\begin{figure}[t!]  \centering
  \includegraphics[width=0.95\columnwidth]{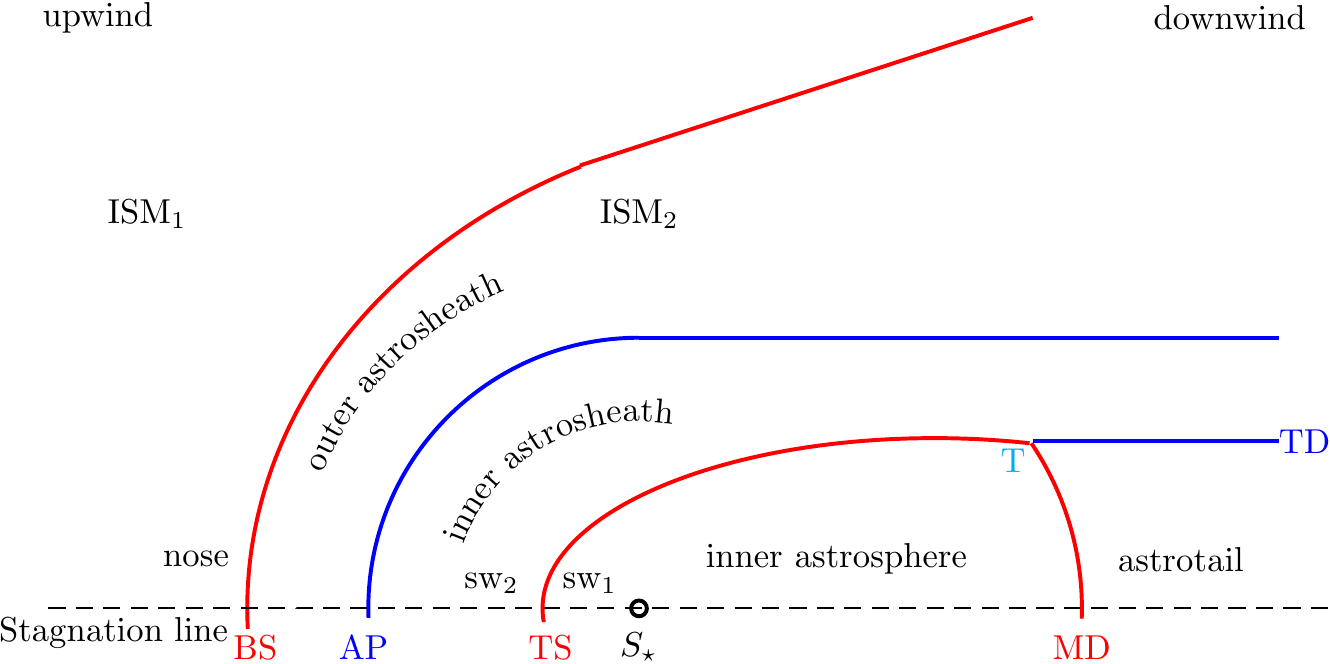}
  \caption{Large-scale structure of an astrosphere. For details
    see text.}
  \label{fig:as}
\end{figure}

The hydrodynamic large-scale structure is sketched in
Fig.~\ref{fig:as}, the notation of which is described below.

The hypersonic stellar wind (Mach numbers $Ma \gg 1$) undergoes a
shock transition to subsonic velocities at the termination shock (TS)
in the inflow direction. Then a tangential discontinuity, the
astropause (AP), is formed between the ISM and the stellar wind, where
the velocity normal to it vanishes: there is no mass transport through
the AP. Other quantities such as the tangential velocity, temperature, and
density are discontinuous, while the thermal pressure is the same on
both sides.  If the relative speed, or the interstellar wind speed as
seen upwind in the rest frame of the star, is supersonic in the ISM, a
bow shock (BS) exists. If the relative speed is subsonic, there will be no BS, see
Table~\ref{tab:3} for the stellar parameters and for the
stellar-centric model distances of the TS, AP, and BS. The region
between the BS and AP is called outer astrosheath, the region between
the AP and TS the inner astrosheath. The AP around the
inflow direction at the stagnation line is sometimes called the nose,
while the region beyond the downwind TS is called the astrotail. The
latter can extend deep into the ISM. The region inside the TS is
called the inner astrosphere.

In the downwind direction, the termination shock forms a triple point,
from which the Mach disk (MD) extends down to the stagnation line; this is
the line through the stagnation point and the star. A tangential
discontinuity (TD) emerges down into the tail. A reflected shock
(not shown here) also extends from the triple point toward the TD.

This is the standard shape of an astrosphere using a single
hydrodynamic fluid \citep{Baranov-etal-1971,Pauls-etal-1995}. For the
large dimensions of O-star astrospheres, cooling operates inside
the outer astrosheath, but usually not in the inner astrosheath.
Cooling is also present beyond the model boundary (see Table~\ref{tab:2}) of the inner
astrosphere, which leads to relative sizes of these regions different
from that of pure hydrodynamic flow.

The astrosphere is always bullet-shaped, which can be seen by the
conservation of momentum:
$\rho_{sw} v_{sw}^{2} + P_{sw} = \rho_{ISM}v_{ISM}^{2} + P_{ISM}$,
where $\rho, v, P$ are the density, velocity, and the thermal pressure of
the stellar wind (subscript $_{sw}$) and the ISM (subscript $_{ISM}$).
The stellar wind pressure is usually neglible inside the TS. For
hypersonic flows this holds true for the thermal ISM pressure in
inflow direction, while for subsonic inflows the thermal pressure
dominates. In the tail direction, only the thermal ISM presssure is
present beyond the TS.  This means that even if the inflow is subsonic,
there is a total pressure asymmetry between upwind and downwind as
long as the ISM velocity does not vanish. Thus on long time and large
spatial scales, a bullet-shaped astrosphere will develop.

\section{Cooling and heating}
\label{sec:1}
As a result of the large dimensions of astrospheres and their outer
astrosheath, the plasma can effectively be cooled by Coulomb
collisions. In this process, an electron in an atom (molecule)
will be excited, and after returning to a lower energy state, the
re-emitted photon will carry away the energy
\citep[][ and references therein]{Sutherland-Dopita-1993}. This can
lead to an effective cooling of the shocked ISM (ISM$_{2}$ in
Fig.~\ref{fig:as}).

Several cooling functions are discussed in the literature
\citep[e.g.,\
][]{Rosner-etal-1978,Mellema-Lundqvist-2002,Townsend-2009,Schure-etal-2009,Reitberger-etal-2014}.
The differences in the cooling functions are caused by different
abundances and different levels of approximation. Because we do not
know the abundances in the ISM surrounding \lamC, we use here the
analytic representation by \citet{Siewert-etal-2004}, which lies in
between all the other cooling functions mentioned above, and can
therefore be considered as a useful principal representation.

Mainly the hot shocked ISM is affected, the
stellar wind gas is not. The reason for this is that the number densities in
the shocked ISM are higher by a factor of 3.6, which is the compression
ratio between the shocked and unshocked ISM, and that it is by a
factor 10$^{2}$ hotter, that is,\ $T_{ISM,shocked}\approx 2.5\cdot
10^{5}$\,K.
In this region most of the mentioned cooling functions yield similar
values. Inside the inner astrosheath, that is, the region between the AP
and the TS, the number density is on the order of
$n=10^{-3}$\,cm$^{-3}$ and the velocities are on the order of
$v=600$\,km/s, and thus the cooling length scale, depending on $v$ and
$n,$ becomes huge and is neither important there nor inside the TS,
see below.

The heating length scale only depends on the number density $n$ and
 also increases to scales much larger than the distances inside the
AP. Therefore, as a result of the huge ram pressure of the stellar wind inside
the TS, we can, from a dynamical point of view, safely neglect heating
and cooling, because it does not influence the adiabatic expansion of
the stellar wind. Moreover, the thermal pressure is negligible
compared with the ram
pressure in the inner astrosphere.  With the help of the
momentum equation, we can uniquely determine the shocked thermal
pressure, which dominates in the inner astrosheath.

In the following we always assume quasi-neutrality
$n_{e}=n_{p}\equiv n$, where $n_{e},n_{p}$ are the electron and proton
number density, respectively. Furthermore, we neglect the contribution
of heavy ions.  In Table~\ref{tab:3} we summarize the parameters of
\lamC derived from observations and the characteristic
distances of its model astrosphere.  The parallax translates into a
tangential velocity of 41.1\,km/s, which together with the radial
velocity of 75.1\,km/s gives a total speed of 85.5\,km/s. Our
hydrodynamic model is three-dimensional (for later use), and in view
of the uncertainties of the ISM state, we have chosen a set of
parameters as given in Table~\ref{tab:2}. The standard procedure in
modeling is to choose one axis as the inflow direction (here the y
axis). As long as the ISM is homogeneous, with a vanishing magnetic field,
and the stellar wind is spherically symmetric, the astrosphere is
symmetric around the inflow direction.

The derived inner boundary conditions for the model are estimated from
the mass-loss rate and the terminal velocity and taken at 0.03\,pc.
They are presented in Table 3 together with those of the ISM.

\begin{table}[t!]
  \centering
  \begin{tabular}{l|l}
    \hline
    $\dot{M}$ [$M_{\odot}$/yr] & $1.5\cdot10^{-6}$\\
    terminal velocity [km/s] & 2500\\
    spectral type & O6If(n)p \\
    distance [pc] & 649$^{+112}_{-63}$\\
    radial velocity (redshift) [km/s] & -75.1\\
    parallax [mas/yr] & -7.46 and -11.09\\
    \hline
  \end{tabular}
  \caption{Parameters for \lamC
    \citep{van-Leeuuwen-2007}. }
  \label{tab:3}
\end{table}

\begin{table}[t!]\label{tab:2}
 \begin{tabular}{l|rr}
&   \multicolumn{1}{c}{ at 0.03\,pc} & \multicolumn{1}{c}{ISM}\\
   \hline
$n$ [part./cm$^{-3}$]   &    6  &  11\\
$v$ [km/s]              &  2500 &  80\\
$T$ [K]                 & 917  & 9000\\
    $R_{TS}$ [pc] & 0.66 \\
    $R_{AP}$ [pc] & 0.93 \\
    $R_{BS}$ [pc] & 1.1 \\
 \end{tabular}
 \caption{Stellar and interstellar boundary conditions. The
   stellar-centric distances of the termination shock  $R_{TS}$, the
   astropause $R_{AP}$, and the bow shock $R_{BS}$ are 
   from the model described here. The stellar wind parameters are
   taken from \citet{Henrichs-Sudnik-2013}. As interstellar
   temperature we use the diffuse ionized medium 
   temperature derived by \citet{Reynolds-etal-1999}. The density
   value is estimated from integrating the HI slice at the postion 
   and the velocities expected for the distance of $\lambda$ Cephei based
   on data of the Canadian Galactic Plane Survey  \citep{Taylor-etal-2003}.}
\end{table}

\begin{table*}[t!]
  \centering\label{tab:1}
\begin{tabular}{cccccccc}
       & $T$\,[K]   &$v$\,[km/s] & $n$\,[\#/cm$^{-3}$] &
       $L_{cool,s}$ & $\tau_{cool,s}$ & $L_{heat,s}$
       & $\tau_{heat,s}$ \\
\hline
ISM$_{2}$:   & $10^{6}$  &20 & 40   & 25\,kAU$\approx$ 0.12\,pc & 6\,kyr &$ 248\,\mathrm{AU}\approx 10^{-3}\,\mathrm{pc}$  & 58\,kyr \\
SW$_{2}$: & $10^{6}$ & $10^{3}$ & $10^{-3}$ & 51\,GAU
$\approx$ 250\,kpc &  242 Myr &  1241\,GAU $\approx$ 6\,Mpc & 5.8 \,Gyr\\
\hline
       &    & &  & $L_{cool,h}$ & $\tau_{cool,h}$ & $L_{heat,h}$                        & $\tau_{heat,h}$ \\
ISM$_{1}$   & 10$^{4}$ & 80 & 10 & 56\,kAU $\approx $ 0.3 pc & 3.3 kyr &2.8\,MAU $\approx$  14 pc &168 kyr\\
\hline
\end{tabular}
  \caption{Some characteristic numbers using the cooling function from
    \citet{Siewert-etal-2004}.}
\end{table*}

As stated above, the cooling acts differently in the subsonic regions
from the way it does in the supersonic regions because in the former the thermal
pressure $P$ is dominant, while in the latter it is the ram pressure
$\rho v^{2}/2,$ where $\rho$ and $v$ are the mass density (mainly
protons, but helium or other elements may contribute to the mass
density) and bulk speed, respectively.  While in the following we
discuss only protons, we can add, for example, helium, which leads to
partial densities, temperatures, and pressures for which the
approximations made below can be separated. We are interested in the
shortest characteristic length, which is in the subsonic case
inversely proportional to the number densities, and thus the protons
dominate.

In the hypersonic case, the estimation below can differ by up to 40\%
when including helium. This would then also require including helium
as a new species in the Euler energy equations
\citep[including different species, see ][]{Scherer-etal-2014}, which
would violate our assumption of a single fluid. The interaction of
other species concerning the energy loss by Coulomb collisions is,
however, already included in the cooling functions, so that we can
continue with the single-fluid equations consisting of protons including
the cooling term for a first analysis.

From the stationary energy equation we obtain in the subsonic region with
the assumption $ n m_{p} v^{2}/2 \ll P $:
\begin{eqnarray}
  \label{eq:c1}
\div \left(\frac{\gamma}{\gamma -1}P + \frac{1}{2} n
  m_{p} v^{2}  \right) \vec{v} 
 &\approx& \frac{\gamma P v}{(\gamma-1)L_{cool,s}}
,\\\nonumber &=& \frac{5 n k T v} {L_{cool,s}} = -n^{2} \Lambda(T)
\end{eqnarray}
$\text{where }    \gamma =5/3$ is the adiabatic index, $m_{p}$ the proton mass,
$P=2n k T$, $k$ is the Boltzmann constant, $\vec{v}, T$ the bulk
velocity and temperature of the plasma flow, and $\Lambda$ the cooling
function. Taking the absolute values in Eq.~(\ref{eq:c1}) and
replacing $\nabla$ by the inverse subsonic cooling length,
$L_{cool,s}$ can be estimated as follows:
\begin{eqnarray}
  \label{eq:c2}
  L_{cool,s} \approx \dfrac{\gamma P v}{(\gamma-1) n^{2} \Lambda(T)} =
  \dfrac{5 k T v}{n\Lambda(T)}
,\end{eqnarray}
and the subsonic  cooling time $\tau_{cool,s}$: 
\begin{eqnarray}
  \label{eq:c3}
  \tau_{cool,s} = \dfrac{L_{cool,s}}{v} = \dfrac{5 k T }{n\Lambda(T)}
.\end{eqnarray}
This cooling time is up to a factor 3 the same as that given in
\citet{Sutherland-Dopita-1993}. 

For supersonic interstellar flows, that is,\
$ \rho_{ism}v^{2}_{ism} / 2 \gg \gamma/(\gamma - 1)P$,  it follows:
\begin{eqnarray}
  \label{eq:c4}
\div \left(\frac{\gamma}{\gamma -1}P + \frac{1}{2} n
  m_{p} v^{2}  \right) \vec{v} \approx {\cal O} \left(\frac{m_{p} nv^{3}}{2 L_{cool,h}}   \right)    
,\end{eqnarray}
and we derive the supersonic (index $h$ for ``hypersonic'' to distinguish
it from the subsonic one) cooling length $L_{cool,h}$ and time
$\tau_{cool,h}$
\begin{eqnarray}
  \label{eq:c5}
  L_{cool,h} \approx \dfrac{m_{p} v^{3}}{2 n \Lambda(T)} \qquad
  \mathrm{and} \qquad\tau_{cool,h}
  =  \dfrac{m_{p} v^{2}}{2 n\Lambda(T)}
.\end{eqnarray}

For the heating function by photo-ionization and some supplemental
heat source we use the approach by \citet{Reynolds-etal-1999}; see
also \citet{Kosinski-Hanasz-2006}. The heating rate for
photo-ionization depending on electron collisions is limited by
recombination and is thus proportional to $n_{e}^{2}$, while
additional heating terms that can include photoelectric heating by
dust, dissipation of turbulence, interactions with cosmic rays are
proportional to $n_{e}$:
\begin{eqnarray}
  \label{eq:h1}
  \Gamma = n^{2} G_{0} + n G_{1} 
,\end{eqnarray}
with the constants $G_{0} = 10^{-24}$\,erg cm$^{3}$ s$^{-1}$ and
$G_{1} = 10^{-25}$\,erg\,s$^{-1}$ \citep{Kosinski-Hanasz-2006}.
Replacing the right-hand side of Eq.~(\ref{eq:c1}) by $\Gamma$, we obtain
\begin{eqnarray}
  \label{eq:h2}
  L_{heat,s} = \dfrac{5 k T  v}{n G_{0} + G_{1}} \qquad \mathrm{and} \qquad
  L_{heat,h} = \dfrac{m_{p} v^{3}}{2 (n G_{0} + G_{1})} 
,\end{eqnarray}
and the heating times $\tau_{heat,s}, \tau_{heat,h}$ by dividing the
respective heating lengths by the speed $v$.

\begin{figure}[t!]
  \center
   \includegraphics[width=0.9\columnwidth]{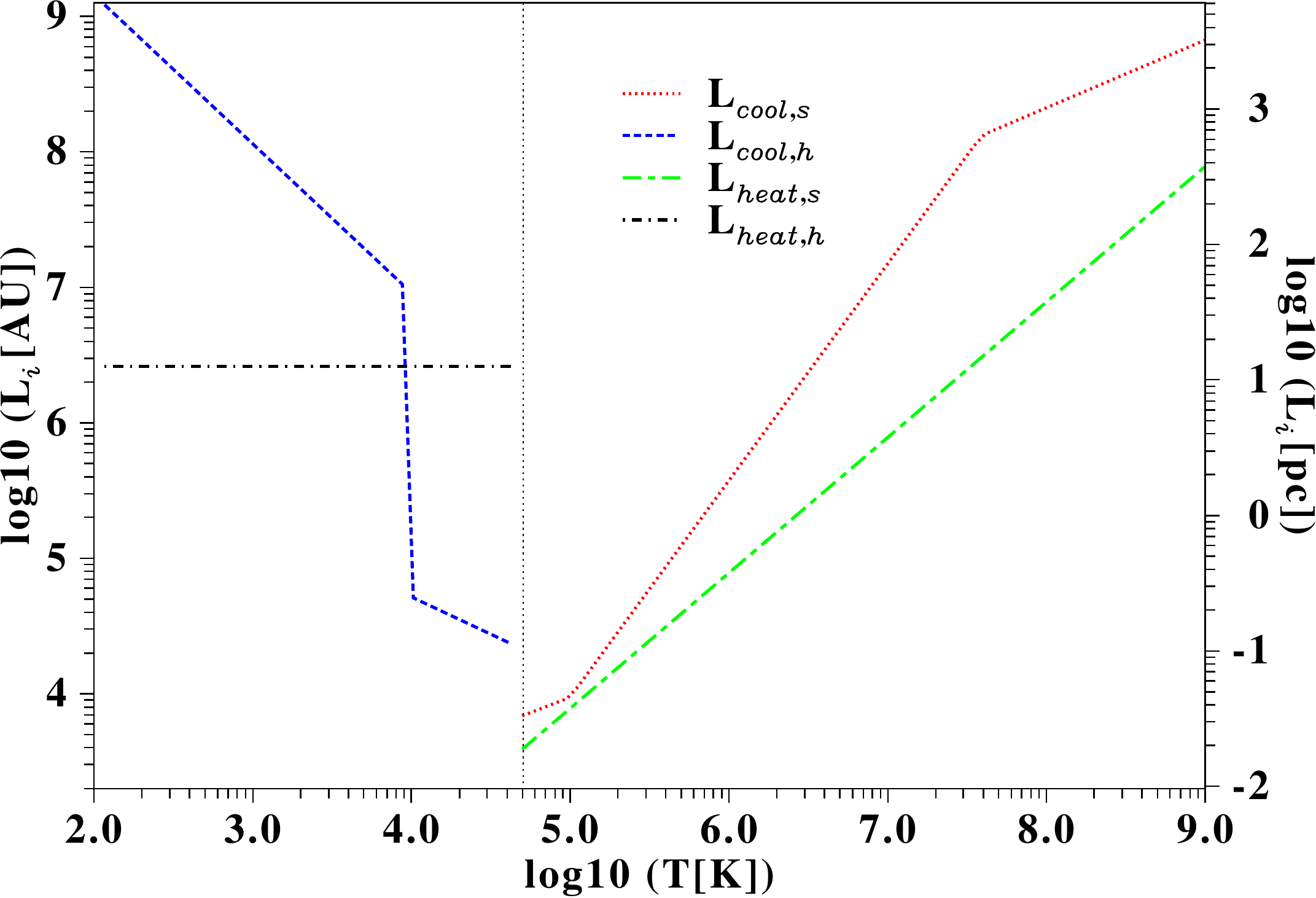}
  \caption{\label{fig:0}Number density and speed are fixed at $n=
     11$\,part/cm$^{3}$ and $v=80$\,km/s. All scales are
   logarithmic. On the horizontal axis the temperature is displayed, while on
   the left vertical axis the characteristic length is shown in AU and
 in parsec on the right y-axis. The black vertical line denotes the
 temperature where the ram pressure equals the thermal pressure.}
\end{figure}

We can estimate the cooling and heating lengths and times for the
shocked ISM (ISM$_{2}$), and analogously for the shocked stellar wind
$SW_{2}$ (see Fig.~\ref{fig:as}).  The results are displayed in
Table~\ref{tab:1} together with the characteristic lengths and times
for the hypersonic ISM (ISM$_{1}$), because it is also cooled. For the
ISM the dependence of the cooling and heating lengths is shown in
Fig.~\ref{fig:0}. The vertical line denotes the temperature at which
the ram pressure equals the thermal pressure. Left of this line the
hypersonic length scales from Eqs.~(\ref{eq:c5}) and~(\ref{eq:h2}) are
displayed, while on right the subsonic scales are shown
(Eqs.~(\ref{eq:c2}) and~(\ref{eq:h2})).

The supersonic parameters are number density $n= 11$\,cm$^{-3}$ and a
speed of $v=80$\,km/s while for the subsonic parameters, the density was
multiplied by the compression ratio $s=3.62$ and the speed was divided
by $s$ using the Rankine-Hugoniot relations.

Figure~\ref{fig:0} shows that in the subsonic case for
temperatures above $\approx 5\cdot10^{5}$\,K the heating scale length is always longer than that for the cooling. Thus cooling is more efficient than heating for the discussed
functions and parameters.  In the supersonic case
($T<5\cdot10^{5}$\,K), the heating scale lengths are only
longer down to temperatures of $\approx 10^{4}$\,K, while for
temperatures below $\approx 4\cdot 10^{2}$\,K the cooling scale length
is more important. Thus, above $\approx 4\cdot 10^{2}$\,K
cooling is more efficient than heating, and below this, heating is important.
We can read from this figure that the length scale for cooling of the
shocked ISM temperature $T\approx 2.5\cdot10^{5}$\,K is $\approx
0.01$\,pc.
From the model we see that the BS to AP distance is 0.17\,pc, which is
more than ten times the cooling length $L_{cool,s}$. This strong
cooling is balanced by the fact that at least at the stagnation line
the shocked ram pressure $1/2 \rho_{2,ISM} v_{2,n,ISM}^{2}$ has to be
converted into thermal pressure toward the astropause, through which
mass flux is zero.

It is also evident from Fig.~\ref{fig:0} that if the temperature falls
below $\approx 4\cdot 10^{4}$\,K, the flow again becomes supersonic.

Different values of the number density $n$ or speed $v$ or different
cooling or heating functions change the characteristic lengths,
but the principle estimates for the dynamics of the flow field remain
the same.

In the stationary case, where no relative speed between the star and
the ISM exists, one should take for the speed $v$ that of the outward-moving shock front.  Moreover, the criterion given by
\citet{Schwarz-etal-1972}, namely that $d\log(\Lambda(T))/d\log(T)>2$
to guarantee thermal stability, is generally not satisfied, and thus
one expects clumping in between the AP and the BS and possibly in the
ISM.  The cooling lengths can be scaled by the speed and the density:
$L_{cool,s}$ does not change as long as $v/n= \const$ This is not
true for the heating length $L_{heat,s}$.

A similar result was also derived by
\citet{van-Marle-Keppens-2011}. In the supersonic ISM the ram pressure
dominates the cooling scales.  It is expected that a magnetic
field will increase the characteristic
length scale due to its additional pressure and thus helps in stabilizing the shock structures
\citep{van-Marle-etal-2014}. The above estimate of the characteristic
cooling length is crucial in determining the resolution required for
numerical models of the large-scale structure of astrospheres. From
Fig.~\ref{fig:0} it is clear that this should be higher than the shortest
characteristic cooling lengths, which are on the order of $>$0.01\,pc
for the numbers given there. Moreover, the cooling of the
surrounding ISM can also be easily be inferred from the above estimate, and
thus it can be approximately determined whether it is cooled during the computation.

\section{Astrosphere model}
\label{sec:2}

\begin{figure}[t!]
\centering 
  \includegraphics[width=0.9\columnwidth]{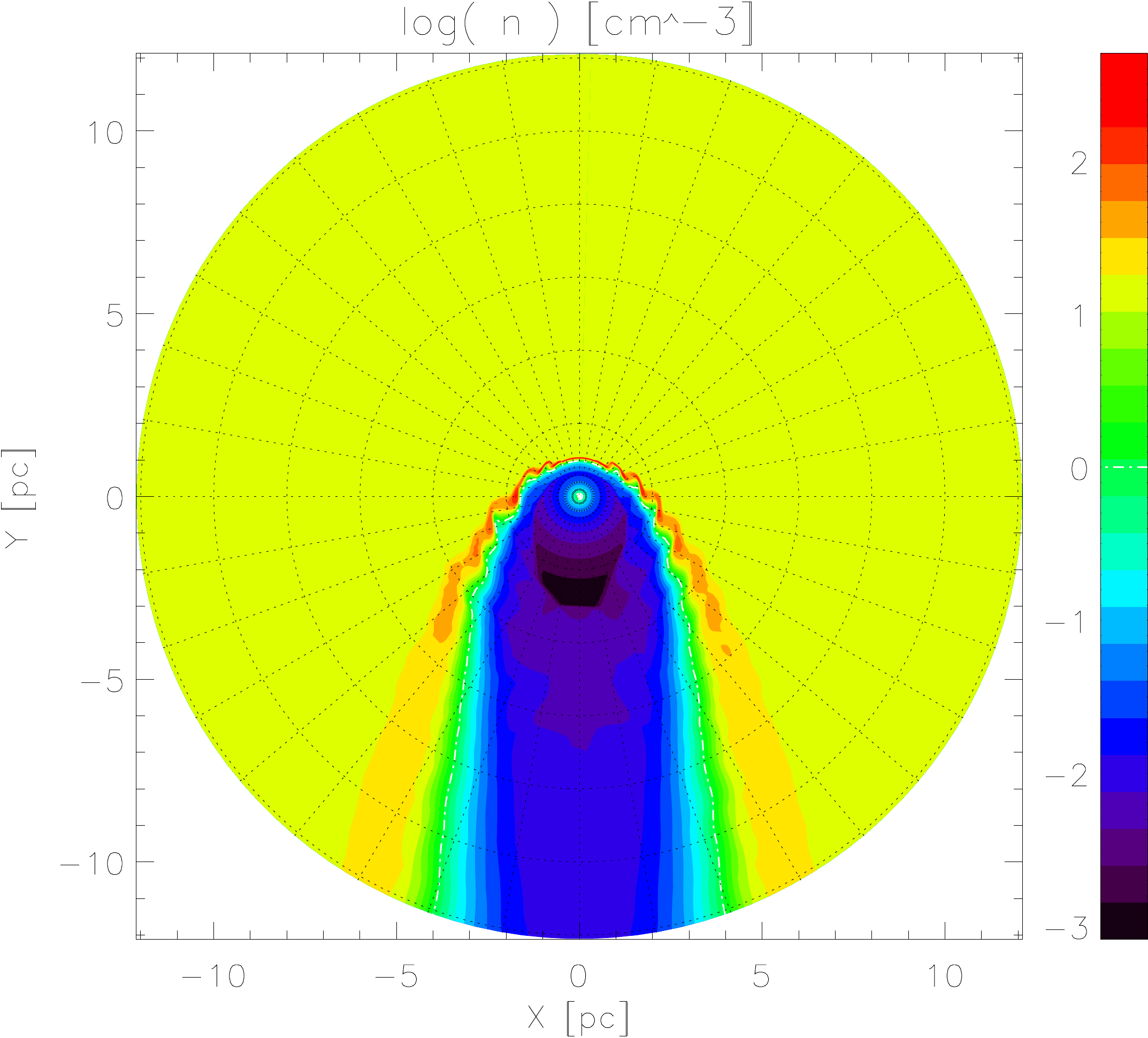}
  \caption{\label{fig:1a}Logarithmic number density for \lamC
    (including heating and cooling). The axes are given in pc, while
    the color bar is a logarithmic scale for the proton number
    density.}
 \end{figure}

The hydrodynamic model for the star \lamC is made in 3D; the third dimension
is needed for future comparison with models including bipolar winds or
magnetic fields. The boundary conditions are given in
Table~\ref{tab:2}.  While the stellar mass loss and the terminal speed
can be determined by observations, the stellar wind temperature as
well as the interstellar parameters are sophisticated guesses.  The
model solves the Euler equations for \lamC using the Cronos MHD code
as described in \citet{Kissmann-etal-2008},
\citet{Kleimann-etal-2009}, and \citet{Wiengarten-etal-2013}. The
results are displayed in Fig.~\ref{fig:1a} for the proton number
density of \lamC. In Fig.~\ref{fig:1a} the wiggles along the AP caused
by the thermal instabilities can be clearly recognized. They are due to
the cooling functions.

This model provides the underlying plasma structure needed as input
for the transport equation discussed below. The model is solved on a
spherical grid with a resolution of 0.005\,pc in radial dimension, and
2$^{\deg}$ and 3$^{\deg}$ in $\vartheta$ and $\varphi$ dimension,
respectively. The large distance of the outer boundary is needed
because during the evolution of the astrosphere it becomes much
broader than shown in Fig.~\ref{fig:1a} and finally shrinks to the
state shown here after ca. 170\,kyr.

\begin{figure}[t!]
\centering 
  \includegraphics[width=0.9\columnwidth]{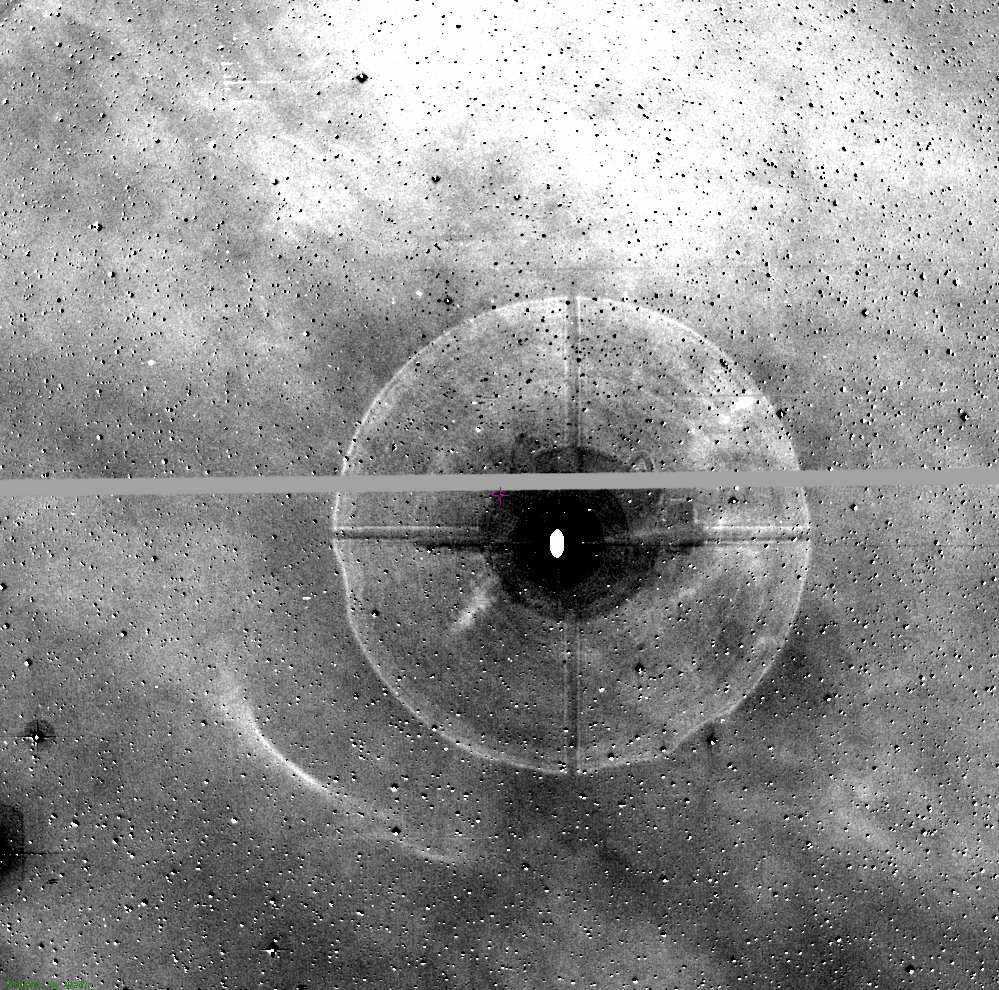}
  \caption{\label{fig:1b} Continuum-corrected and
    slightly spatially filtered H$\alpha$ image of the bow shock
    around $\lambda$ Cephei.  North is up, east is left, the size of
    the plotted field is $22\arcmin \times 22\arcmin$.  The imaging
    data were taken as part of the IPHAS \citep{Drew-etal-2005}
    survey.  A bright and well-defined bow shock nebula is visible
    toward the SE of the star.  The roughly spherical H$\alpha$
    structure centered on $\lambda$ Cephei is an out-of-focus ghost of
    the bright star, the dark inner rings are the equivalent structure
    from the r-filter image used for continuum subtraction.  In addition to
    these purely instrumental effects, several H$\alpha$
    bright features are visible close to the star.  A more detailed
    discussion of the multiwavelength properties of the bow shock and
    the circumstellar environment of this runaway O6 star will be
    presented in an upcoming paper.}
 \end{figure}

\section{Cosmic ray fluxes}
\label{sec:3}

To model the flux of GCRs, the Parker transport equation
\begin{equation}
\frac{\partial f}{\partial t} = -\vec{v} \cdot \nabla f + \nabla \cdot \left( \mathbf{K} \cdot \nabla f \right) + \frac{\nabla \cdot \vec{v}}{3} \frac{\partial f}{\partial \ln P} 
\end{equation}
has been frequently used in the literature
\citep[see][and reference therein]{Potgieter-etal-2001a}. In this
equation, $f$ is the (nearly) isotropic GCR distribution function,
$\vec{v}$ the bulk plasma flow, $\mathbf{K}$ the diffusion tensor
(including, in a 3D geometry, separate components directed parallel and
perpendicular to the mean magnetic field), and $P$ is the particle
rigidity. As an initial approach, spherical symmetry is assumed, so
that the Parker equation reduces to
\begin{equation}
\frac{\partial f(r,t)}{\partial t} = -v \frac{\partial f}{\partial r} + \frac{1}{r^2} \frac{\partial}{\partial r} \left( r^2 \kappa_{rr} \frac{\partial f}{\partial r}  \right) + \frac{P}{3r^2} \frac{\partial }{\partial r} \left( r^2 v \right) \frac{\partial f}{\partial \ln P}
\label{Eq:TPE_1D}
,\end{equation}
where $r$ is radial distance and $\kappa_{rr}$ is the effective radial
diffusion coefficient. In this work, Eq.~(\ref{Eq:TPE_1D}) is solved
by transforming it into the equivalent set of stochastic differential
equations (SDEs)
\begin{eqnarray}
dr &=& \left[ \frac{1}{r^2} \frac{\partial}{\partial r} \left( r^2 \kappa_{rr} \right)  -v \right]dt + \sqrt{2 \kappa_{rr}}  dW ,\\
dP &=& \left[ \frac{P}{3r^2} \frac{\partial}{\partial r} \left( r^2 v \right) \right]dt
,\end{eqnarray}
which is then integrated numerically \citep{Strauss-etal-2011a}, where
the solution of the 1D equation along $r$ can be found in
\citet{Strauss-etal-2011a}.

These equations are coupled to the simulated HD geometry by reading
in the modeled values of $\vec{v}$ and $\nabla \cdot \vec{v}$
(governing energy changes) directly from the HD simulations (along
the stagnation line) and solving for the GCR flux using essentially a
test particle approach. The magnetic field enters the computations via
the diffusion coefficient $\kappa_{rr} = \kappa_{rr}(B)$. In the 1D
scenario, the geometry of $\vec{B}$ does not enter the computations,
although for a azimuthal field (the case inside the TS),
$\kappa_{rr} \approx \kappa_{\perp}$, thus reducing to a diffusion
coefficient perpendicular to the mean field. As a boundary condition
for the GCR flux, a local interstellar spectrum is specified at the
edge of the computational domain. The GCR differential intensity  is
related to the distribution function by $j=P^2f$.

Based on experience gained from modulation studies inside the
heliosphere, $\kappa_{rr}$ can be decomposed into a radial and energy
dependence, $\kappa_{rr} = \kappa_1(r)\kappa_2(P)$, where, for this
study
\begin{equation}
\kappa_1(r) = \left\{ \begin{array}{r}
\kappa_{sw} \hspace{1mm}  \mathrm{if}  \hspace{1mm} r < r_{BS}   \\
\kappa_{ism}   \hspace{1mm}  \mathrm{if}  \hspace{1mm} r \geq r_{BS} \end{array} \right.
,\end{equation}
with $r_{BS}$ the radial position of the BS and where $\kappa_{ism}$
is independent of position and $\kappa_{sw}=\kappa_0
B_0/B$.
$\kappa_0$ is a normalization constant, usually specified near the
inner boundary, and $B_0$ is a constant added for dimensional
consistency. Assuming that the astrospheric magnetic field is about 80
times higher than the heliospheric case \citep{Naze-2013}, a
value of $0.027$ kpc$^2$Myr$^{-1}$ is used, scaling up the
heliospheric diffusion coefficient by about two orders of magnitude
for this astrospheric case. The $B^{-1}$ radial dependence is based on
the results of \citet{Engelbrecht-Burger-2013}, but approximated in such a
fashion because in situ observations are, of course, not
available. The energy dependence of $\kappa_{rr}$ is taken from \citet{Buesching-Potgieter-2008}
\begin{equation}
\kappa_2(P) = \left\{ \begin{array}{c}
\left( \dfrac{P}{P_0} \right)^{+ 0.6} \hspace{1mm}  \mathrm{if}  \hspace{1mm} P > P_0   \\[.5cm]
\left( \dfrac{P}{P_0} \right)^{-0.48}   \hspace{1mm}  \mathrm{if}  \hspace{1mm} P \leq P_0 \end{array} \right.
,\end{equation}
with $P_0 = 4$ GV. Instead of showing $\kappa_{rr}$, we follow the
convention of  showing the corresponding mean free path (mfp),
\begin{equation}
\lambda = \frac{3\kappa_{rr}}{v_p} 
,\end{equation}
where $v_p$ is particle speed. The magnitude of $\kappa_{rr}$ is
changed in the next section to illustrate its effect on the resulting
particle intensities.

Because the mfp used is,  in general, not known, we studied the
behavior of three different values inside \lamC and three different values
of the interstellar mfp.

\subsection{Mean free path}
\begin{figure}[t!]
   \includegraphics[width=.925\columnwidth]{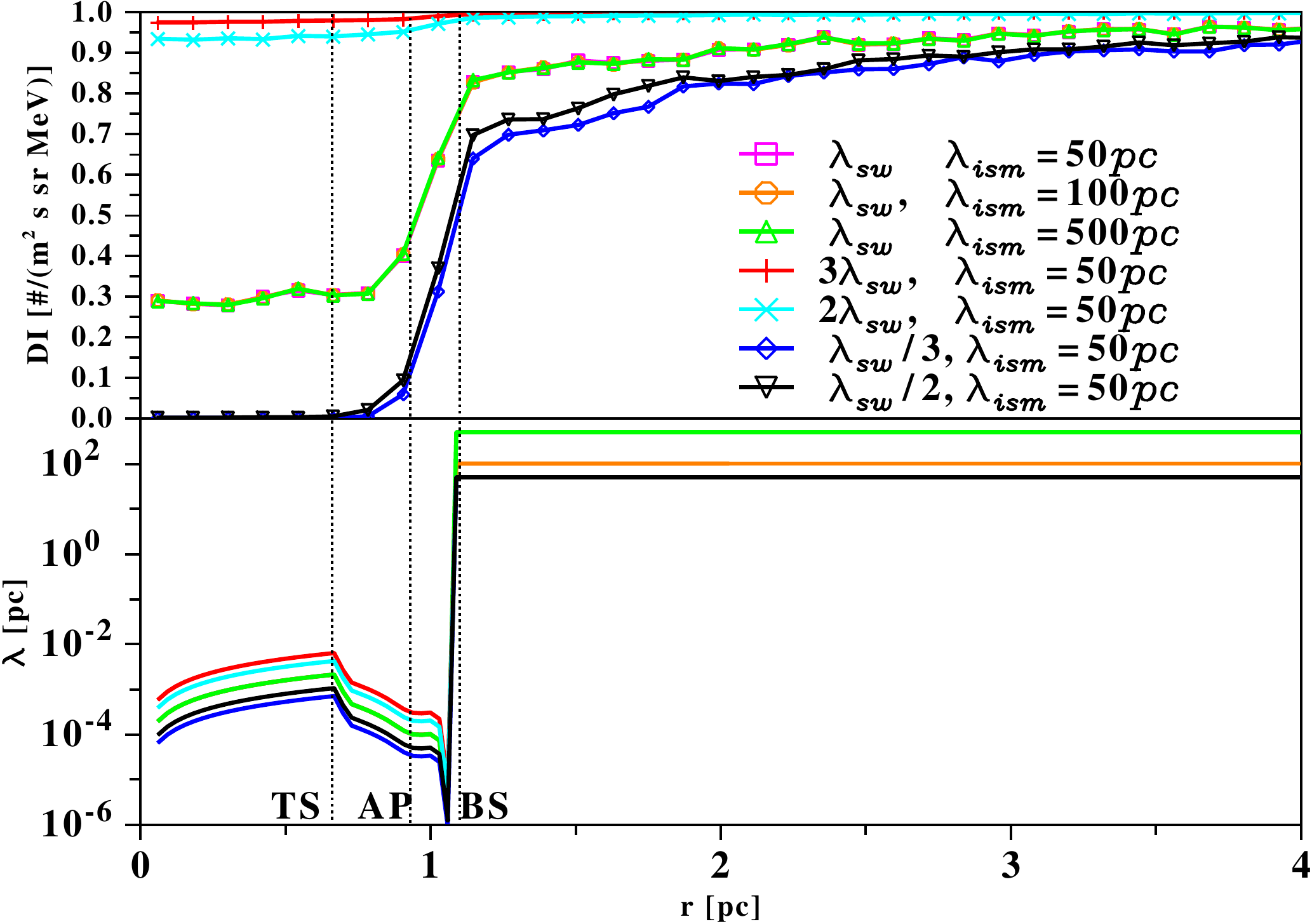}
   \caption{ \label{fig:3} Modulation of 1 GeV particles in \lamC
     along the stagnation line for different stellar and interstellar
     mfp. The mfp are shown in the lower panel and the colors
     correspond to those in the upper panel. The magenta, orange, and
     green curves in the upper panel are identical when normalized to
     their respective values at infinity. The dotted lines show the
     position of the TS, AP, and BS.}
\end{figure}

The structure of the astrosphere can be recognized in the lower part
of Fig.~\ref{fig:3} by the respective jumps in the mfps (from the
right, which is the inflow direction): the BS (1.1\,pc), AP (0.93\,pc)
and TS (0.66\,pc), represented by the vertical dotted lines. This TS
is marked by sharp decrease of the mfp, the AP by a change of the slope,
and finally the BS by the sharp drop at the ISM side. The resulting
differential flux (DI) for 1\,GeV particles along the stagnation line
is displayed in Fig.~\ref{fig:2}. In all cases the
modulation, that is, the decrease in the differential intensity DI,
starts far away in front of the BS, a feature that was discussed by
\citet{Scherer-etal-2011}, \citet{Herbst-etal-2012}, and
\citet{Strauss-etal-2013} for the case of the heliosphere.

This outer astrospheric modulation depends on the ratio between the
parallel and perpendicular diffusion coefficient and
vanishes for high ratios. Thus, for an appropriate choice of
the transport parameters, the modulation of GCRs starts in front of the
astrosphere and then rapidly decreases at the astrospheric BS. It is
also evident that in contrast to the heliosphere, the GCRs are
modulated directly behind the BS: This is due to the cooling effects,
which shrink the region between the AP and BS to 0.17\,pc and are a
barrier-like feature for the cosmic rays
\citep{Potgieter-Langner-2004}.

In the lower part of Fig.~\ref{fig:3} we plot the mfp used in the modeling. The inner mfp ($\lambda_{sw}$) obtained from the model are then
divided or multiplied by a factor two and three to demonstrate its
influence. A short $\lambda_{sw}$ strongly increases
the modulation, and the flux of 1\,GeV particles almost
vanishes inside the AP . Increasing $\lambda_{sw}$ by the same factor leads
to a nearly vanishing modulation. Changing the
interstellar $\lambda_{ism}$ from 50\,pc to 100\,pc
and 500\,pc does not change the DI remarkably. These
variations show the theoretical expected effects. Especially for short
mfp inside the astrosphere the GCR spectra are efficiently cooled. The
20\% modulation in front of the astrosphere is presumably caused by
the sharp drop of the mfp between the AP and the BS.

From Fig.~\ref{fig:3} we can also see that most of the modulation --
up to a factor 5 or even more -- occurs between the BS and the AP
(depending on the transport parameters). Thus, because astrospheres
are large volumes surrounding the star, the GCR spectrum can be cooled
and, especially, small-scale anisotropies in the galactic CRF can be
observed far away from large astrospheres. This prediction will be
improved by further modeling and studying the distribution of hot
stars in the Galaxy.

At the BS, the mfp falls to quite low values, rises sharply at the AP,
and finally jumps again at the TS. In the inner astrosheath the mfp
increase continuously toward the shock because we assumed that the
magnetic field is inversely proportional to the speed $v$ as for the
Parker spiral field in the heliosphere
\citep{Potgieter-etal-2001a}. Thus at the stagnation line, the speed
must vanish at the AP because this is a tangential discontinuity with
no mass flow through it. Inside the TS, the magnetic field increases
with $r$ in the direction to the star, and the mfp decreases. Because of
the strong compression of the outer astrosheath, the mfp diminishes to
an even lower value, which then forms a barrier for the cosmic ray
transport.

This qualitative behavior is theoretically well understood and
confirmed by observations in the heliosphere, except for the
additional modulation in the outer astrosheath. The latter differs
because of the cooling function. Thus, the cooling function has a
direct influence on the cosmic ray transport.

\subsection{Energy dependence}
\begin{figure}[t!]
   \includegraphics[width=0.95\columnwidth]{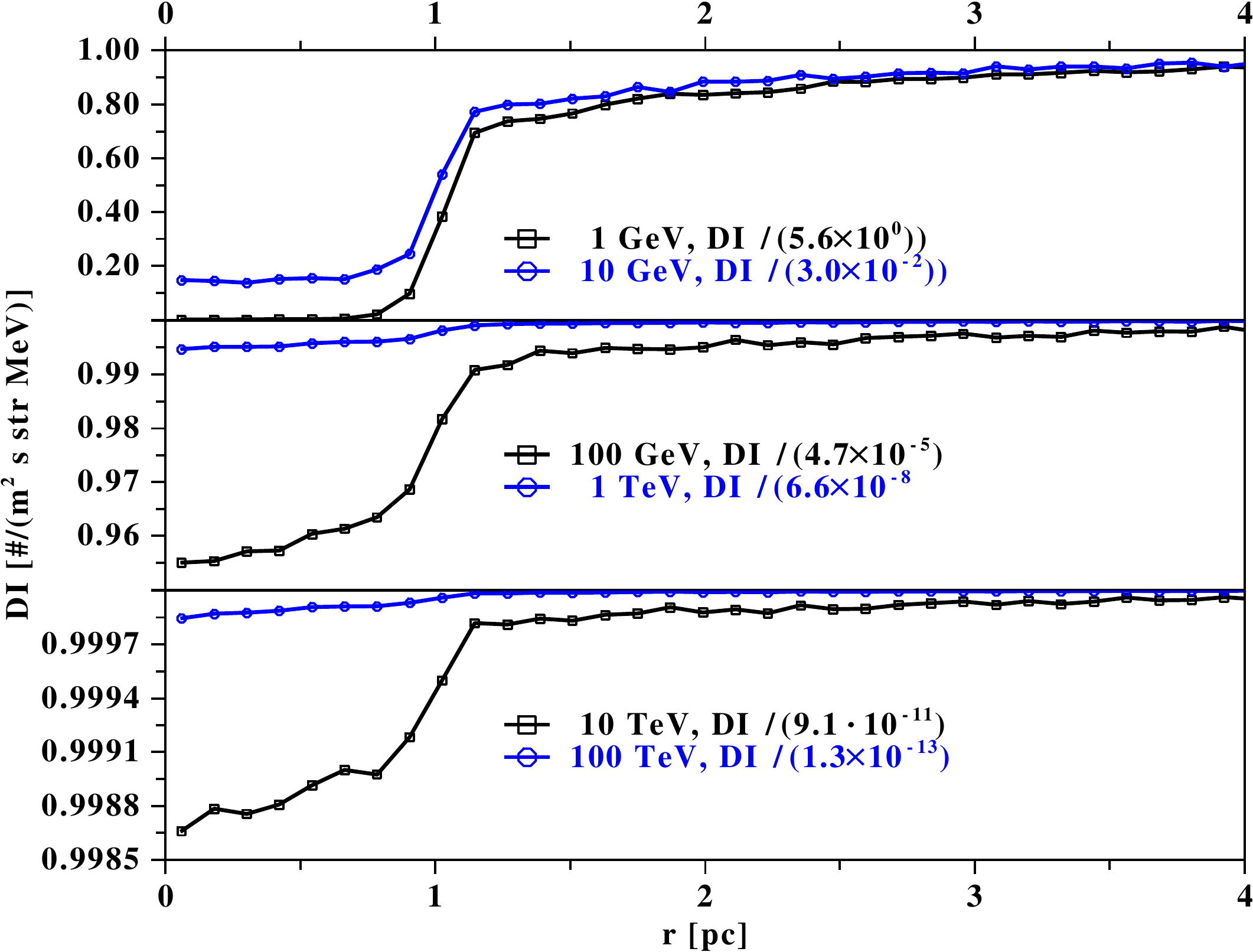}
   \caption{\label{fig:2} The modulation of particles with different
     energies. In the upper panel are the normalized differential
     particle intensities in \unit{ $part./(m^{2} s\, sr\, MeV)$} for
     energies of 1 and 10\,GeV are shown (in black and blue,
     respectively).  The y-axes are in a linear scale with different
     values for all three panels. In the middle panel the DI for
     particles at 100\,GeV and 1\,TeV are displayed (black and blue).
     In the lower panel the DIs for 10 and 100\,TeV particles are
     shown.  All Dis are normalized to their corresponding
     interstellar DI and are decrease towards the star. }
\end{figure}

In Fig.~\ref{fig:2} we study the behavior of the GCR flux at different
energies.  The 1\,GeV particles are modulated by a
factor of almost five, while for higher energies this factor becomes
smaller, as expected. But even for the highest energies of
10.. 100\,TeV, a small modulation of a few per mille
(1 per mille = 0.1\%) outside of the BS is visible.

In Fig.~\ref{fig:2} we plot the DIs for different energies, from 1\,GeV to
100\,TeV in logarithmic steps of ten. In the upper panel the
lower energies (form 1 to 10\,GeV) are strongly modulated, while for
the 10\,GeV and 1\,TeV particles the DI is only diminished by a few
percent (middle panel). The lower panel shows the modulation of the 10
and 100\,TeV particles, which is in the per mille range or even much
lower for the 100\,TeV particles, which are hardly affected. All
energies are modulated ahead of the BS, and thus the modulation volume
is larger than that of the astrosphere.  The DI of the TeV particles
can be increased by diminishing the mfp inside the astrosphere or can
become less by increasing the mfp; this is not shown in Fig.~\ref{fig:2}.
The mfp is based on that in the heliosphere, and the modulations
displayed in Fig.~\ref{fig:2} are based on it.

\subsection{Observational evidence}
Small-scale modulations of a few per mille in the TeV range are
observed with, for example, the IceCube experiment
\citep[e.g.,\ ][ and references therein]{Abbasi-etal-2012}.
In particular, their Fig.~5 shows a resolution of 3$^{o}$ , which
cannot resolve the astrospheres of \lamC with angular diameters
0.2$^{o}$. Nevertheless, the pixels show variations that may
be attributed to local GCR fluctuations, provided the statistics in
the pixels is large enough and the astrospheres are close
enough. Because runaway O/B stars or OB associations are quite
frequent in the Galaxy \citep{Huthoff-Kaper-2002}, their local
disturbance of the GCR spectrum can explain small-scale variations in
the GCR flux. \lamC was only used as an example, but the above holds true
in general for all large-scale astrospheres.

Our simulations may help to understand these variations. The variation
in the DIs can be increased or decreased by choosing a smaller
or larger mfp
inside the astrosphere, therefore a better estimate of their magnetic
fields is needed, which can help to understand their turbulence
levels.

For the higher energies a modulation outside the astrosphere can
also be observed, where the ``astrosphere of influence'' is roughly twice
the BS distance for bullet-shaped astrospheres like that of \lamC. This holds
true in the direction of the inflowing ISM, while it is more
complicated in other directions. These details will be explored in
future. We studied here only protons, but because the transport
coefficients depend on rigidity, the behavior of other species can
roughly be estimated.\ For helium, for instance,  a similar behavior
is expected as for the protons shifted by a factor of two in Fig.~\ref{fig:2}. At
these high energies the rigidity has roughly the same value as the
particle energy. This means that multiplying the rigidity by a factor of 2 is
the same as doing it in the particle energy. Therefore, helium is slightly less modulated when passing through astrospheres than the protons.

If there were a few large astrospheres (or stellar wind bubbles) in
the direction toward an observer, the GCR spectrum would be slightly cooler than in other directions. This can contribute to the
understanding of the small-scale anisotropies present in the IceCube
data \citep{Abbasi-etal-2012}.

The parameter set for the supersonic wind and ISM are chosen to be
close to the observed values, but for modeling purposes we worked
with rounded values. In addition, for the transport model we
took the turbulence
levels based on that of the heliosphere. Both of these
aspects need further attention, but the principal effect remains: large
stellar astrospheres can affect the local interstellar cosmic ray
spectra.

\section{Conclusion}

Based on our models, we studied the transport of GCRs in an
astrosphere. We have shown that even ahead of an astrosphere, where
the mfp is still undisturbed, a cooling of the GCR spectra occurs
because particles are trapped by scattering into the astrosphere, in
which they can be effectively cooled. The ``astrosphere of
influence'' around the modeled astrospheres is roughly twice its
hydrodynamic dimension. Thus, stellar wind bubbles (astrospheres) can
cool the Galactic spectrum, and small anisotropies are expected
in
the direction to an observer.

As a result of its effect in compressing the outer
astrosheath, the cooling function directly influences the GCR modulation in this
region. Additionally, the cooling and heating functions require high
resolutions for global models because of their characteristic length
scales. They also give a rough estimate whether the surrounding ISM is
also affected during the calculations.

The modulation in astrospheres affects particles up to 100\,TeV. This
can help to understand the anisotropies on small angular scales, for
example, by the IceCube experiment, among others.  The angular extent of
\lamC is too small to be resolved by these experiments because of its
large distance ($\approx$650\,pc). Nevertheless, nearby hot stars or
stellar associations can have an larger angular extent and may
possibly be observed.  In our simulations we saw an effective cooling
of the GCR spectrum. Thus, the conclusion is that small-scale cosmic
ray anisotropies may be explained by the modulation in such huge
cavities.

Because the model is fully 3D, the modulation or other parameters along a line of sight
toward Earth can in principle be theoretically determined. Here we demonstrated that large astrospheres can
modulate the Galactic cosmic ray spectrum quite significantly, and no homogeneous spectrum over all directions
can be expected. To
calculate the modulation of GCRs along a line of sight or at Earth
requires more sophisticated methods than discussed here, but
this is being
developed.

\begin{acknowledgements} 
  KS, HF, JK, and TWi are grateful to the
  \emph{Deut\-sche For\-schungs\-ge\-mein\-schaft, DFG\/} funding the
  projects FI706/15-1 and SCHE334/10-1. DB is supported
  by the DFG Research Unit FOR 1254.  This work is also based on the
  research supported in part by the South African NRF. Any opinion, finding, and
  conclusion or recommendation expressed in this material is that of
  the authors, and the NRF does not accept any liability in this
  regard. KS, HF and RDS appreciate discussions at the team meeting
  ``Heliosheath Processes and Structure of the Heliopause: Modeling
  Energetic Particles, Cosmic Rays, and Magnetic Fields'' supported
  by the International Space Science Institute in Bern, Switzerland.  

  This paper makes use of data obtained as part of the INT Photometric
  H$\alpha$ Survey of the Northern Galactic Plane (IPHAS,
  www.iphas.org) carried out at the Isaac Newton Telescope (INT). The
  INT is operated on the island of La Palma by the Isaac Newton Group
  in the Spanish Observatorio del Roque de los Muchachos of the
  Instituto de Astrofisica de Canarias. All IPHAS data are processed
  by the Cambridge Astronomical Survey Unit, at the Institute of
  Astronomy in Cambridge. The bandmerged DR2 catalogue was assembled
  at the Centre for Astrophysics Research, University of
  Hertfordshire, supported by STFC grant ST/J001333/1.
  \\
\end{acknowledgements}


\end{document}